\documentclass{optica-article}

\journal{opticajournal} % for journals or Optica Open

\articletype{Research Article}
\usepackage{booktabs}
\usepackage[table]{xcolor}
\usepackage{colortbl} 
\usepackage{xcolor}
\usepackage{array} 
\usepackage{graphicx}
\usepackage{lineno}
\begin{document}

\title{Efficient multiplexed quantum memory with high dimensional orbital angular momentum states in cold atoms}

\author{Xin Yang, \authormark{1,$\dag$} Chengyuan Wang,\authormark{1,$\dag$} Jinwen Wang,\authormark{1} Mingtao cao,\authormark{2,*} Yun Chen,\authormark{3} Hong Chang,\authormark{2} Ruifang Dong,\authormark{2} Shougang Zhang,\authormark{2} Dong Wei,\authormark{1} Pei Zhang,\authormark{1,*} Fuli Li,\authormark{1} Hong Gao,\authormark{1,*}}

\address{\authormark{1}Ministry of Education Key Laboratory for Nonequilibrium Synthesis and Modulation of Condensed Matter, Shaanxi Province Key Laboratory of Quantum Information and Quantum Optoelectronic Devices, School of Physics, Xi’an Jiaotong University, Xi’an 710049, China\\
\authormark{2}Key Laboratory of Time and Frequency Primary Standards, National Time Service Center, Chinese Academy of Sciences, Xi'an 710600, China\\
\authormark{3}Department of Physics, Huzhou University, Huzhou 313000, China\\
\authormark{$\dag$}The authors contributed equally to this work}
\email{\authormark{*}mingtaocao@ntsc.ac.cn,\authormark{*}zhang.pei@mail.xjtu.edu.cn,\authormark{*}honggao@xjtu.edu.cn} %% email address is required; see note below about the corresponding author designation

% use {asbstract*} to suppress the copyright line. Copyright information will be added in production

\begin{abstract*} 
Quantum memory plays a pivotal role in the construction of quantum repeaters, which are essential devices for establishing long-distance quantum communications and large-scale quantum networks. To boost information capacity and signal processing speed, the implementation of high-efficiency multiplexed quantum memories is essential for the development of multiplexed quantum repeaters. In this work, we experimentally demonstrate an efficient multiplexed quantum memory by consolidating photons carrying high-dimensional orbital angular momentum (OAM) state from 4 individual channels into an elongated cold $^{87}$Rb atomic ensemble. Benefiting from the cold atomic ensemble with high optical depth, we obtain a storage efficiency exceeding 70\% for the 4-channel multiplexed beam. The storage fidelities surpass 83\% when all channels are operated in a 4-dimensional Hilbert space, which is far above the classical benchmark and validates the quantum functionality of our memory system. The achieved high-efficiency OAM multiplexed quantum memory opens up an avenue for efficient quantum information processing over multiple parallel channels, promising significant advances in the field of quantum communication and networking.

\end{abstract*}

%%%%%%%%%%%%%%%%%%%%%%%%%%  body  %%%%%%%%%%%%%%%%%%%%%%%%%%
\section{Introduction}

Quantum repeater-based quantum networks facilitate the long-distance distribution of quantum states between quantum nodes, which is essential for realizing distributed quantum computation, metropolitan-scale quantum communication, and large-scale quantum sensing\cite{lago2021telecom,liu2024creation,knaut2024entanglement}. In recent years, a significant breakthroughs in quantum networks based on remote entanglement between adjacent quantum nodes have been successfully demonstrated across various physical platforms including atomic ensembles\cite{liu2021heralded,liu2024creation}, single atoms\cite{hofmann2012heralded,van2022entangling}, diamond nitrogen-vacancy centers\cite{humphreys2018deterministic}, trapped ions\cite{krutyanskiy2023entanglement} and solid-state system\cite{knaut2024entanglement}. One of the key considerations in advancing the practical implementation of quantum networks is the limited entanglement swapping rate between adjacent quantum nodes \cite{wehner2018quantum,pirandola2016physics}. 
%For the practical implementation of quantum networks, a critical challenge that must be addressed is the low entanglement swapping rate between adjacent quantum nodes\cite{wehner2018quantum,pirandola2016physics}.
Quantum memories\cite{lvovsky2009optical}, which enable the storage and retrieval of quantum states with fidelity surpassing any classical device, are beneficial for synchronizing probabilistic single photons and effectively increasing the entanglement swapping rate. By temporarily storing quantum states, quantum memories mitigate the inherent randomness in photon transmission, thereby enhancing the overall efficiency of entanglement distribution. To optimize this process, enhancing the storage efficiency and multimode storage capability of quantum memories represents a highly effective strategy for improving the entanglement distribution efficiency
%The storage efficiency and multi-mode capacity are two critical factors that determines the entanglement distribution efficiency
\cite{PhysRevLett.98.190503}. For instance, a 1\% increase in storage efficiency can shorten 7-18\% entanglement distribution time, depending on the quantum repeater protocol \cite{RevModPhys.83.33}. Specifically, storing N modes simultaneously can reduce the entanglement distribution time by a factor of N \cite{RevModPhys.83.33,PhysRevLett.98.190503}.

%Quantum memories\cite{lvovsky2009optical}, which enable the storage and retrieval of quantum states with fidelity surpassing any classical device, are beneficial for synchronizing probabilistic single photons and effectively  increasing the entanglement swapping rate within quantum repeater nodes \cite{pu2021experimental,Su_2021,PhysRevLett.131.033601,bhaskar2020experimental,luo2022postselected,liu2021heralded,Lei:23}. 

%Quantum repeater, which comprises quantum channels and quantum nodes for the transmission and manipulation of quantum states, is pivotal in the construction of large-scale quantum networks. Quantum memories\cite{lvovsky2009optical}, which enable the storage and retrieval of quantum states with fidelity surpassing any classical device, are beneficial for synchronizing probabilistic single photons and effectively increasing the entanglement swapping rate within quantum repeater nodes \cite{pu2021experimental,Su_2021,PhysRevLett.131.033601,bhaskar2020experimental,luo2022postselected,liu2021heralded,Lei:23}.

Recently, an amount of effort has been dedicated to increasing these two parameters by adopting various storage schemes. Utilizing the electromagnetically induced transparency (EIT) protocol, storage efficiencies of 85\% for 2-dimensional polarization qubits\cite{wang2019efficient} and entangled state\cite{cao2020efficient} have been achieved in cold atom ensemble. To increase the mode number, Dong \emph{et al.} encode photons in 25-dimensional perfect optical vortex (POV) modes and attain a storage efficiency of 58\%\cite{PhysRevLett.131.240801}. However, the POV mode faces challenges related to propagation stability and collection efficiency limitations\cite{Pinnell:19}. Another approach to increasing data capacity is multiplexed quantum memory, which can be implemented using multiple degrees of freedom (DoFs) of the photons, such as spatial \cite{Lan:09,parniak2017wavevector,pu2017experimental,lipka2021massively,PhysRevLett.124.240504,wang2021efficient}, temporal \cite{PhysRevLett.124.210504,PhysRevLett.118.210501,PhysRevX.7.021028,Wang202300825}, spectral \cite{PhysRevLett.113.053603,saglamyurek2016multiplexed,seri2019quantum}, or any combination thereof \cite{yang2018multiplexed}. One prominent spatial DoF utilized for multiplexing and transmitting multiple data streams is the orbital angular momentum (OAM) \cite{erhard2018twisted}, which finds wide-ranging applications in free-space optical communications \cite{wang2012terabit}, fiber-optic communications \cite{bozinovic2013terabit,liu2020multidimensional}, and quantum communication \cite{liu2020orbital}. However, challenges arise when mapping large transverse size multiplexed beams into mediums with restricted optical depth and sectional dimensions, resulting in very low storage efficiencies in multiplexed quantum memory schemes \cite{Lan:09,parniak2017wavevector,pu2017experimental,yang2018multiplexed,jiang2019experimental,parigi2015storage}. Despite advancements in OAM quantum memory for higher dimensions \cite{nicolas2014quantum,PhysRevLett.114.050502,ding2019broad,zhang2016experimental} and longer storage times \cite{PhysRevLett.129.193601}, the storage efficiencies remain less than 40\%. Since 50\% storage efficiency is the threshold for surpassing the limit of quantum non-cloning without post-selection and some error correction schemes in linear optics quantum computation, the development of highly efficient and multiplexing-compatible quantum memories remains a significant challenge for practical applications.

In this paper, we demonstrate an experimental realization of high-efficiency OAM-multiplexed quantum memory using a cold atomic ensemble. By utilizing the compressed magnetic optical trap (CMOT), spatial dark line, and Zeeman optical pumping \cite{zhang2012dark}, we obtain a cigar-shaped cold atomic cloud with high atomic density and low ground state decoherence rate. Additionally, we develop a compact system capable of generating a 4-channel OAM-multiplexed beam, with each channel carrying a 4-dimensional OAM state. This beam is then Fourier transformed into the spatial frequency domain, enabling us to fully map it into the cold atoms and ensuring a large effective light-matter interaction volume for achieving high storage efficiency. By employing these technologies, we achieve a storage efficiency exceeding 70\% for the multiplexed beam. The storage fidelities of the retrieved states in 4-dimensional Hilbert space surpass 83\%, much higher than the classical benchmark. The storage efficiency reported here is a huge improvement compared to previous works on multiplexing quantum memory, and our scheme is well compatible with state-of-the-art quantum repeater protocols \cite{liu2021heralded,liu2024creation} and can dramatically increase the entanglement swapping rate by virtue of the large multi-mode capacity with high storage efficiency. 

\section{Experimental setup and results}
\begin{figure*}[!htbp]
  \centering
  \includegraphics[width=5.0in]{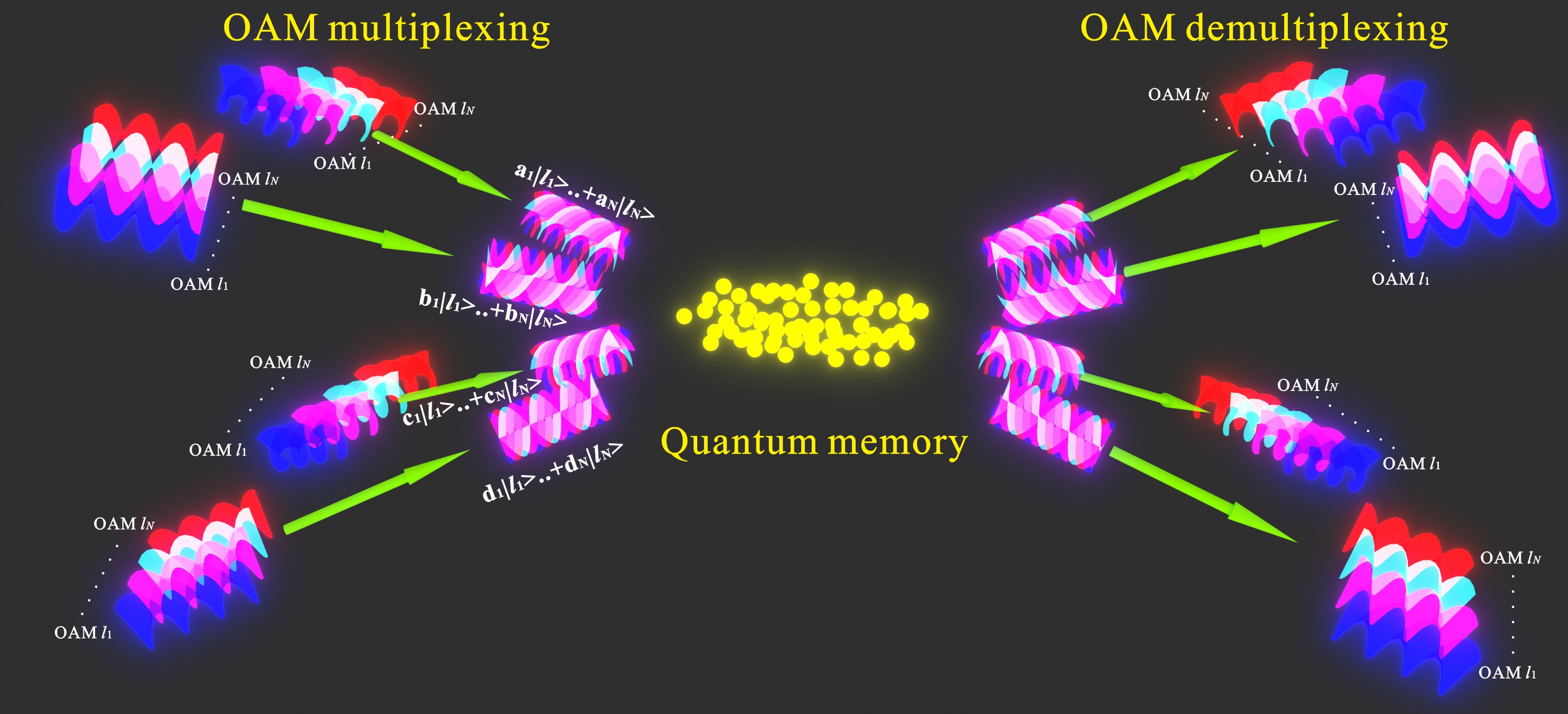}
\caption{Schematic of OAM multiplexed storage. The quantum state in each channel is a superposition of N different OAM states. For instance, the high-dimensional quantum state in the first channel can be written as $|\psi_{input} \rangle = 1\slash\sqrt{N}\sum_{j=1}^{N}\mathrm{a}_j|l_j\rangle$, where $l$ is the OAM quantum number, $\mathrm{a}$ is the weight coefficient. Other channels are encoded in the same way.}\label{fig:0}
\end{figure*}

\textbf{ OAM multiplexed storage architecture.} Figure~\ref{fig:0} presents the schematic diagram of OAM multiplexed storage. In this approach, multiple parallel channels can independently encode photonic information in the high-dimensional OAM space, which is simultaneously projected into the atomic medium for quantum memory. After a programmable time delay, the multiplexed beam is released from the medium, and different channels are separated from each other to achieve demultiplexing. In our proposed scheme, we utilize 4 parallel channels to encode the OAM state, with each channel encompassing a 4-dimensional OAM superposition state. Therefore, we are able to obtain a total of 4 × 4 = 16 modes. This multiplexed beam is subsequently mapped into a cigar-shaped cold atomic cloud with high atomic density to realize high-efficiency quantum memory.

\begin{figure*}[!htbp]
  \centering
  \includegraphics[width=5in]{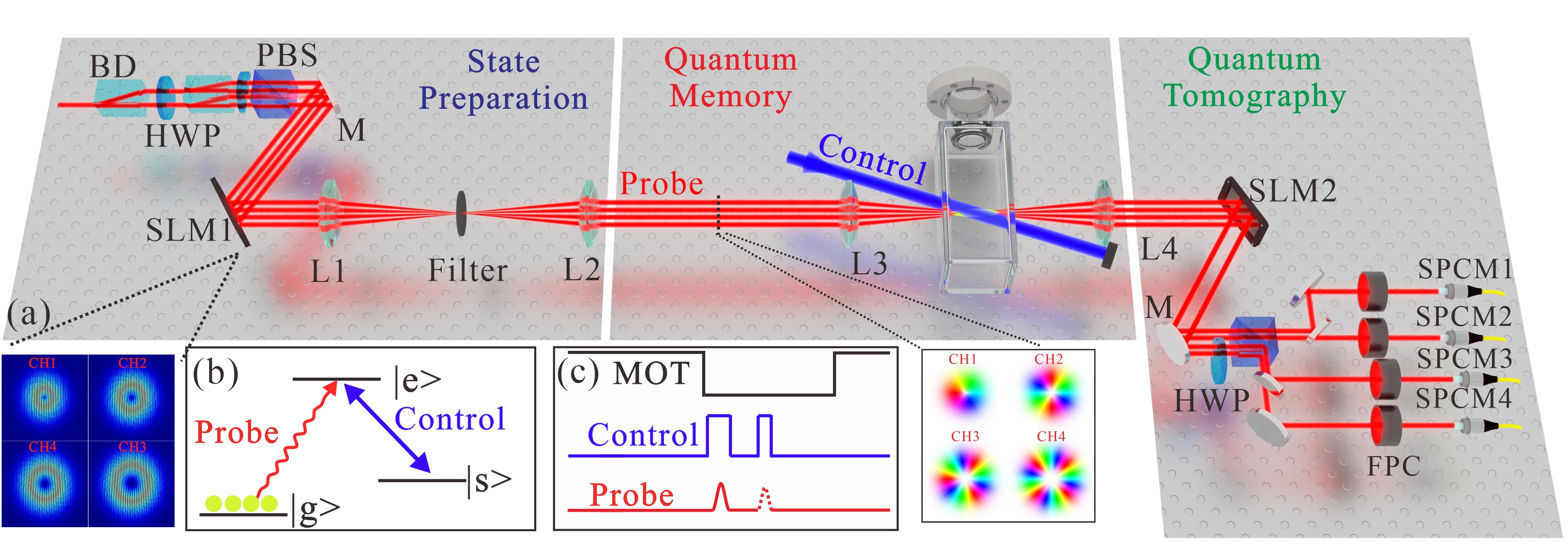}
\caption{EIT quantum memory for OAM multiplexed beam. (a) Architecture of the experimental setup. An elongated cold $^{87} \mathrm{Rb}$ atomic ensemble obtained from a two-dimensional (2D) magneto-optical trap (MOT) acts as the storage medium. The probe beam is modulated into a 4-channel OAM multiplexed beam via two half-wave plates (HWPs), two beam displacers (BDs), a polarization beam splitter (PBS), and a spatial light modulator (SLM). Then it is focused by a lens (L$_3$) and mapped into the atoms for storage by switching on and off the control beam. Another SLM combined with an HWP, a PBS, Fabry–Perot cavities (FPC) filter and single-mode fibers (SMFs) compose the demultiplexing and state projection measurement system. (b) The simplified energy level structure of the quantum memory based on EIT. $\left|g\right\rangle$ and $\left|s\right\rangle$ correspond to $\left|5 \mathrm{S}_{1 / 2}, F=1\right\rangle$ and $\left|5 \mathrm{S}_{1 / 2}, F=2\right\rangle$, which are the two hyperfine ground states of $^{87} \mathrm{Rb}$ $D_{1}$-line, while $\left|e\right\rangle$ is the excited state $\left|5 \mathrm{P}_{1 / 2}, \mathrm{F}^{\prime}=2\right\rangle$. (c) The experimental sequence including the MOT, control and probe field temporal profile.}\label{fig:1}
\end{figure*}

\begin{figure*}[!htbp]
  \centering
  \includegraphics[width=5in]{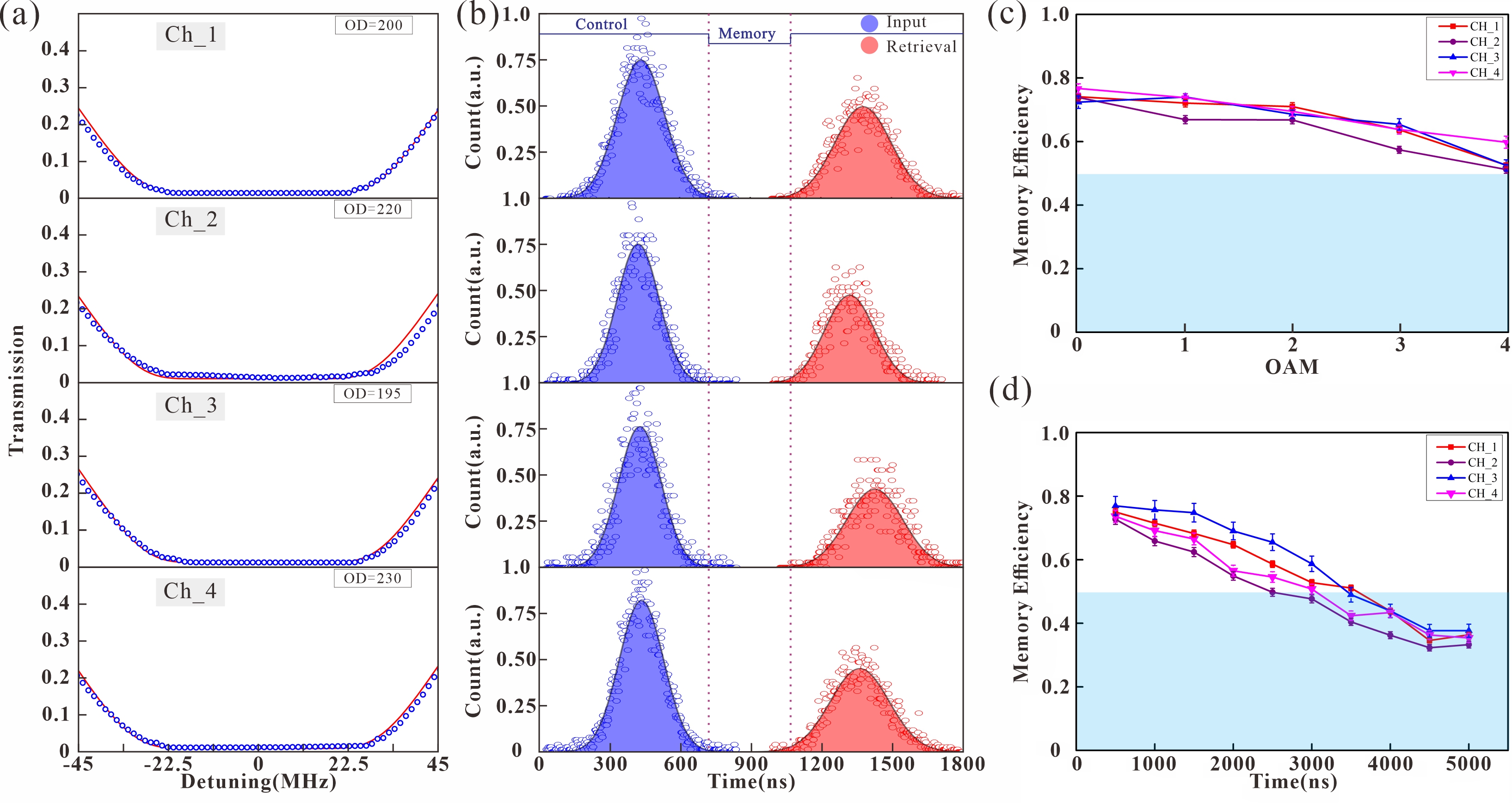}
\caption{Storage performance of the 4-channel multiplexed beam. (a) The optical depth (OD) of different channels. The blue circles are experiment results and the red lines are theoretical fitting. (b) The temporal waveforms of the input probe pulses and the retrieval signals after 500 ns storage time. (c) The storage efficiencies of the channels carrying different topological charges. (d) Variation of storage efficiency with storage time for Gaussian modes in different channels. The blue areas in (c) and (d) represent values below the classical threshold of 50\%.}\label{fig:2}
\end{figure*}

\noindent\textbf{Experimental configuration.} A large OD is a prerequisite to obtain a high storage efficiency \cite{cho2016highly,hsiao2018highly}. This requires the multiplexed beam to interact with as many atoms as possible. To this end, we construct a compact system that can generate an OAM-multiplexed beam with a transverse size smaller than the cross-section of the cold atoms, ensuring that the beam is fully coincident with the long axis of the cigar-shaped cold atomic ensemble. This system is composed of a half-wave plate (HWP), two beam displacers (BDs), a polarization beam splitter (PBS), a spatial light modulator (SLM), and a lens. As seen in Fig.~\ref{fig:1}, a 45$^\circ$-polarized Gaussian beam is divided into two parallel beams with the same power by a BD. A HWP, a PBS, and the other BD can further divide two beams into four horizontally polarized beams (the distance between two adjacent beams is 3.5 mm). These beams are then directed into a Spatial Light Modulator (SLM), which is divided into four segments, each containing a hologram designed to convert the reflected beams into various OAM modes. In this way, an OAM-multiplexed beam comprising 4 channels is obtained. This beam is Fourier transformed into the spatial frequency domain by a lens (L3), ensuring that all channels are fully encompassed by the atomic ensemble to achieve a high OD. The generated OAM multiplexed beams carrying different topological charges and the cross-sectional profiles of the cold atoms can be seen in Supplementary Note for details. After the MOT, the multiplexed beam is collimated by another lens (L4) with the same focal length and is injected into another SLM for mode transformation. In the end, we use an HWP and a PBS to separately couple the 4 channels into single-mode fibers (SMFs) to characterize the output states. The OD measurement outcomes are displayed in Fig.~\ref{fig:2}(a), demonstrating similar OD values of approximately 200 across the 4 channels. This ensures a high and consistent storage efficiency across all channels.

%This requires the multiplexed beam to have a small transverse size to interact with as many atoms as possible since the atomic cloud has a limited cross-section area. To this end, we construct a compact system to generate an OAM-multiplexed optical field suitable for cold atom storage. transverse size of the multiplexed beam to be smaller than the cross-section of the storage medium. The multiplexed beam is then focused at the center of the MOT by a lens with a focal length of 500 mm.

In the following, we perform OAM-multiplexed quantum memory based on the electromagnetically induced transparency (EIT) protocol, which enables the conversion of the probe photon into atomic collective excitation based on the quantum interference between auxiliary control and probe beams interact with three-energy atomic system\cite{liu2001observation,choi2008mapping,vernaz2018highly}. As shown in Fig.~\ref{fig:2}(b), the frequency of the probe beam from an external cavity diode laser (ECDL) is tuned to resonance of the $\left|5 \mathrm{S}_{1 / 2}, \mathrm{F}=1, m_{F}=1\right\rangle \rightarrow\left|5 \mathrm{P}_{1 / 2}, \mathrm{F}^{\prime}=2, m_{F}=2\right\rangle$ transition. A control beam with a waist of 4 mm is injected into the MOT with an angle of 1$^{\circ}$ relative to the probe beam. This beam is phased-locked to the probe beam and resonates to the $\left|5 \mathrm{S}_{1 / 2}, \mathrm{F}=2, m_{F}=1\right\rangle \rightarrow\left|5 \mathrm{P}_{1 / 2}, \mathrm{F}^{\prime}=2, m_{F}=2\right\rangle$ transition to eliminate the photon switching effect \cite{hsiao2018highly}, which can prevent the excess four-wave mixing noise and decrease the ground state decoherence rate. The power of the control beam is 60 mW, which is optimized to ensure minimum probe pulse leakage for the 4 channels during the storage process. To match the spectral-temporal properties of the EIT memory, the probe pulse is modulated to a Gaussian-shaped waveform with a temporal length of 500 ns. The probe beam is turned on 600 $\mu$s after the magnetic coil is turned off to reduce the ground-state decoherence rate induced by the inhomogeneous magnetic field. Once the probe pulse entirely enters the atomic ensemble, the control beam is adiabatically turned off, transferring the probe beam into long-lived atomic collective excitations. After a 500 ns delay, the control beam is switched on to convert the collective excitations into photonic modes. 

%The time sequence of the storage process is given in Fig.~\ref{fig:1}(b).

%%%%%%%%%%%%%%%%%%%%%%%%%%%%%%%%

\noindent\textbf{OAM multiplexed quantum memory.} 
Firstly we characterize the storage performance by storing 4 Gaussian modes. The input pulse is attenuated to contain 3 photons per pulse (measured when the MOT is turned off) and the accumulation time for each measurement is 1000 s. Fig.~\ref{fig:2}(b) gives the experimentally measured temporal waveforms of the input probe pulses and the released signals after 500 ns storage time.  
The storage efficiencies, determined from the ratio of the input pulse areas to the corresponding output pulse areas, are measured to be 74.1 $\pm$ 1.4\%, 76.8 $\pm$ 1.1\%, 72.4 $\pm$ 1.3\%, and 74.4 $\pm$ 0.9\% for the four channels.

Encoding the multiplexed beam with high-dimensional OAM states can provide a larger information capacity. So we also measure the storage efficiencies of the multiplexed beam carrying high-order modes. To this end, all channels are encoded with the same OAM number and vary from $l=0$ to $l=4$. The results are shown in Fig.~\ref{fig:2}(c). We can see that for the four channels, the average storage efficiency decreases from 75\% for $l=0$ to 50\% for $l=4$. This is due to the fact that the beam waist for an OAM mode follows the relationship of $\omega_{l}=\sqrt{l+1}\omega_{0}$, where $\omega_{0}$ is the beam waist for $l=0$. As the atomic density of MOT gradually decreases from the center to the edge region, a higher-order OAM mode experiences a smaller OD and results in lower storage efficiency. More details can be found in the supplementary material. 

We also measure the storage efficiency as a function of storage time, as shown in Fig.~\ref{fig:2}(d). All the channels suffer non-exponential decays of the storage efficiencies against the storage time. This mainly originates from the ground state dephasing induced by the inhomogeneous residual magnetic field \cite{PhysRevLett.111.240503} and atomic thermal motion \cite{PhysRevA.93.063819}. 

In the following, we investigate the high-dimensional storage capability of our system. All the channels are operated in 4-dimensional Hilbert spaces and the input states are firstly encoded to
$| \psi_{input} \rangle = 1\slash\sqrt{4}\sum_{j=1}^{4}|l_j\rangle = 1\slash\sqrt{4}(|\mathrm{-2}\rangle+|\mathrm{-1}\rangle+|\mathrm{1}\rangle+|\mathrm{2}\rangle)$. The storage efficiencies for these qudit states are 71.2 $\pm$ 1.3\%, 65.9 $\pm$ 1.4\%, 70.7 $\pm$ 0.7\%, and 71.4 $\pm$ 0.7\%. The storage efficiency discrepancy among different channels mainly originates from two aspects: the different intersection angles between the control beam and each channel, as well as the different beam waists of each channel. These two aspects lead to an inequable effective interaction area between the control beam and each channel, eventually resulting in different EIT transparency efficiencies.

\begin{figure}[!htbp]
  \centering
  \includegraphics[width=5in]{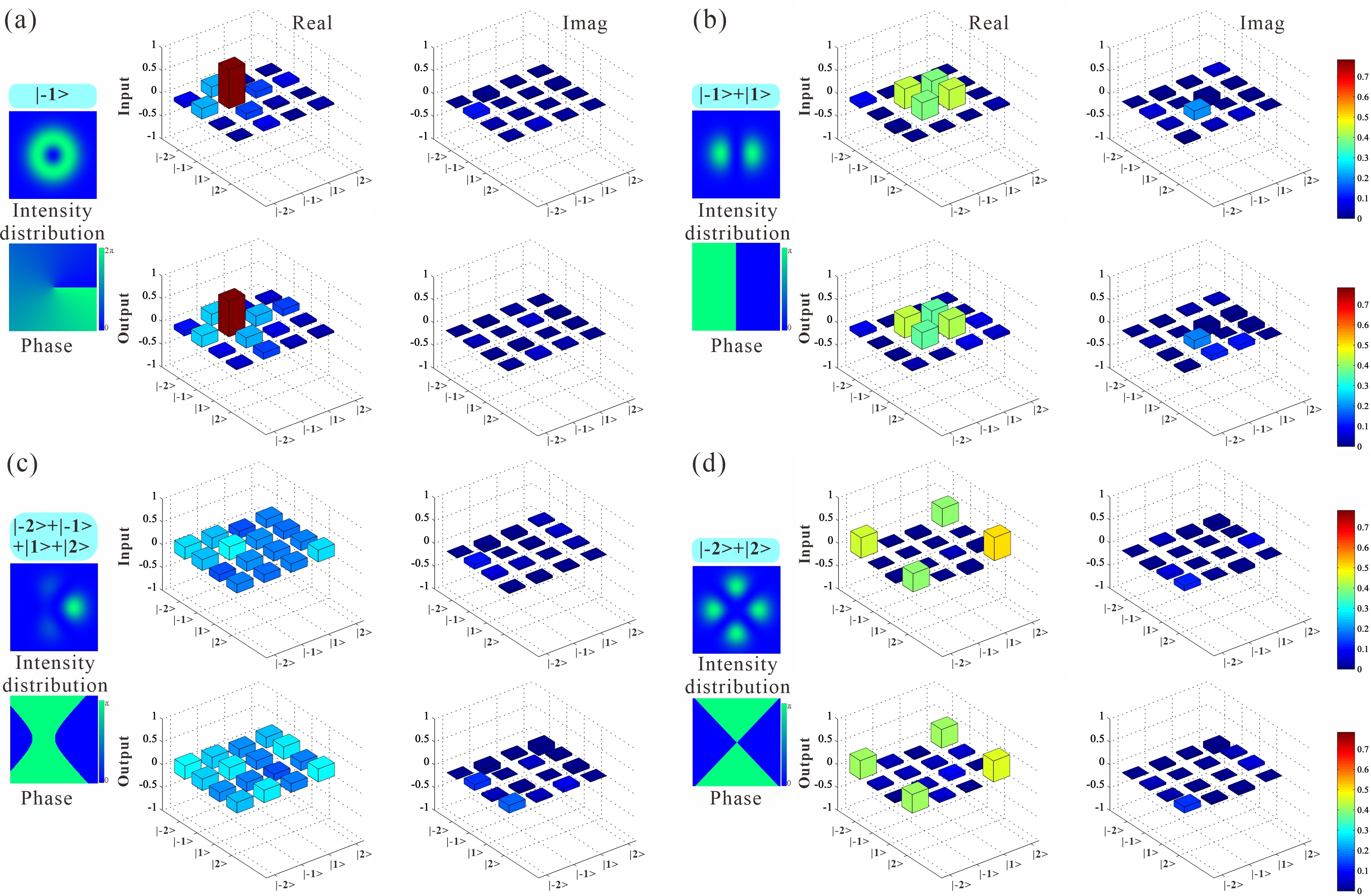}
\caption{The reconstructed density matrices of the 4 channels before (input) and after (output) storage. The input states in (a)-(d) are $|\mathrm{-1}\rangle$, $1\slash\sqrt{2}(|\mathrm{-1}\rangle+|\mathrm{1}\rangle)$, $1\slash\sqrt{4}(|\mathrm{-2}\rangle+|\mathrm{-1}\rangle+|\mathrm{1}\rangle+|\mathrm{2}\rangle)$, and $1\slash\sqrt{2}(|\mathrm{-2}\rangle+|\mathrm{2}\rangle$), respectively.}\label{fig:3}
\end{figure}

To demonstrate the capability of our system to store arbitrary dimensional states and operate in the quantum region, the probe beam is attenuated to contain a mean photon number of 0.5 per pulse and the input states are changed to 4 different states: $|\mathrm{-1}\rangle$, $1\slash\sqrt{2}(|\mathrm{-1}\rangle+|\mathrm{1}\rangle)$, $1\slash\sqrt{4}(|\mathrm{-2}\rangle+|\mathrm{-1}\rangle+|\mathrm{1}\rangle+|\mathrm{2}\rangle)$, and $1\slash\sqrt{2}(|\mathrm{-2}\rangle+|\mathrm{2}\rangle$). The input and retrieved signals are projected into 16 mutual unbiased bases via the SLM and SMFs to construct the density matrices, and the results are shown in Fig.~\ref{fig:3}. The storage fidelity of the output states with the input ones could be evaluated as $\mathcal{F}_{\text {storage}}=[\operatorname{Tr}[\sqrt{\sqrt{\hat{\rho}_{out}} \hat{\rho}_{in} \sqrt{\hat{\rho}_{out}}}]]^{2}$, where $\hat{\rho}_{in(out)}$ is the density matrix of the input (output) state. The storage fidelities for these states are $92.8\%\pm0.2\%$, $96.3\%\pm0.5\%$, $83.2\%\pm0.3\%$, and $85.7\%\pm0.3\%$, respectively. The storage efficiencies and storage fidelities for 4 individual channels as shown in table \ref{table1}. The maximum achievable classical fidelity for a 4-dimensional qudit state with a mean photon number of 0.5 is 43.7\%. All of our results are far above this classical boundary, confirming that our system can faithfully work in the quantum region. The imperfect input density matrices primarily result from the non-orthogonality of different OAM modes induced by perturbations in the optical components.

\begin{table}[htbp]
\caption{ \text{Quantum memory for high-dimensional OAM states from  4 individual channels}}
  \centering
  \resizebox{\linewidth}{!}{
  \begin{tabular}{cccccccc}
    \toprule
     \multicolumn{2}{c}{\textbf{Channel 1}} & \multicolumn{2}{c}{\textbf{Channel 2}} & \multicolumn{2}{c}{\textbf{Channel 3}} & \multicolumn{2}{c}{\textbf{Channel 4}} \\ 
    \cmidrule(lr){1-2} \cmidrule(lr){3-4} \cmidrule(lr){5-6} \cmidrule(lr){7-8} \\
    \text{Efficiency} & \text{Fidelity} & \text{Efficiency} & \text{Fidelity} & \text{Efficiency} & \text{Fidelity} & \text{Efficiency} & \text{Fidelity} \\
   \midrule
     74.1 ± 1.4\% & 92.8 ± 0.2\% & 76.8 ± 1.1\% & 96.3 ± 0.5\% & 72.4 ± 1.3\% & 83.2 ± 0.3\% & 74.4 ± 0.9\%  & 85.7 ± 0.3\% \\
    \bottomrule
  \end{tabular}
  }
  \label{table1}
\end{table}

\section{Discussion}
Our experiment realizes high-efficiency quantum memory of a 4-channel OAM multiplexed beam in laser-cooled $^{87}$Rb atoms. With a well-designed system, we generate an OAM-multiplexed beam with a sectional dimension smaller than that of the cigar-shaped cold atomic ensemble. This guarantees a large OD to achieve a high storage efficiency for the multiplexed beam. We obtain a storage efficiency above 70\% and storage fidelity above 83\% for the 4-channel multiplexed beam carrying different OAM qubits. The presented efficiently multiplexed quantum memory could find applications in constructing large-scale repeater-based quantum networks\cite{PhysRevA.80.022339,Wei202100219} and distributed quantum computing\cite{Oh202300007}. 

In our scheme, the input state is $|\psi_{input} \rangle = 1\slash\sqrt{N}\sum_{j=1}^{N}|l_j\rangle$ while the output state after storage can be written as $|\psi_{output} \rangle = 1\slash\sqrt{\sum_{j=1}^{N}\eta_{j}^{2}}(\eta_1  |l_1\rangle+\cdots+\eta_N|l_N\rangle)$. To achieve a high storage fidelity, the storage efficiency $\eta$ for different OAM modes should be similar to guarantee the weight of each component in the superposition state does not change after retrieval. As can be seen in Fig.~\ref{fig:2}(c), the storage efficiency for $l=\pm$1 and $l=\pm$2 are similar. When the OAM order is above 2, the waist of the multiplexed beam becomes comparable to the cross-sectional size of the cold atoms, resulting in a rapid decline in storage efficiency. Hence we only encode each path with a 4-dimensional OAM superposition state for storage. Further improvement in storage efficiency and capacity could be achieved by optimizing the transverse size of the multiplexed beam and enlarging the cross-section dimension of the atomic ensemble, as detailed in the supplementary material.

The crosstalk between the four channels is negligible owning to the large separation distance between each mode. Benefited by the cold atoms with benign coherence and low thermal velocity, the crosstalk between the orthogonal spatial modes within each channel is also very small. This is revealed in the reconstructed density matrixes, where the components other than the eigenmodes are nearly non-existent.

%The main reason to encode each path with a 4-dimensional OAM superposition with the highest order of $l=\pm$2 is due to that the storage efficiency for $l=\pm$1 and $l=\pm$2 are similar (see Fig.~\ref{fig:2}(c)), which guarantee the weight of different orders in the superposition state do not change after storage. When the OAM order is above 2, the waist of the multiplexed beam become comparable to the cross-section size of the cold atoms, which results in a rapid decline in storage efficiency. 
%Further improvement the in storage efficiency and capacity could be achieved by reducing the transverse size of the multiplexed beam using nanostructured films such as metalens array \cite{li2020metalens} and increasing the density, cross sectional size and length of the atomic ensemble. $|\psi_{output} \rangle = 1\slash\sqrt{\sum_{j=1}^{N}\eta_{j}^{2}}\sum_{j=1}^{N}\eta_j|l_j\rangle$.   

\section{Methods}

The cigar-shaped cold $^{87}$Rb atomic ensemble with a temperature of 200 $\mu$K is obtained from a magneto-optical trap (MOT) system. The experiment runs 50 ms per cycle. The atoms within a vacuum chamber are cooled by 3 pairs of cooling beams (with a total power of 300 mW) and two pairs of rectangular trapping coils (operating with a coil current of 3.5 A) within the first 40 ms. To further increase the atomic density, we employ the CMOT technique and ramp the coils' current from 3.5 A to 8 A to compress the atomic cloud within the subsequent 8 ms. A copper line with a diameter of 1 mm is implemented to block the center region of the repump beam, which can create a dark line along the longitude direction of the cold atoms. In such a way, all atoms are populated into $\left|5 \mathrm{S}_{1 / 2}, \mathrm{F}=1\right\rangle$ within the dark line region to prevent the radiation trapping loss and atom re-heating. Subsequently, an additional optical pumping beam with a power of 6 mW and a diameter of 5 mm is applied for 30 $\mu$s at the end of the compressed MOT stage. This beam pumps all the atoms into the Zeeman state $\left|5 \mathrm{S}_{1 / 2}, \mathrm{F}=1,m_{F}=1\right\rangle$. The atomic density can be substantially increased using the techniques above. In the last 1 ms, we measure the OD of the atomic ensemble and conduct the quantum memory experiment. The relationship between probe transmittance $T$ and OD can be expressed as $T=\mathrm{exp}({\rm Im}(\frac{\mathrm{OD\delta\gamma_{ge}}}{2\delta(\delta+i\gamma_{ge}/2)}))$, where $\delta$ is the detuning of the probe beam, $\gamma_{ge}=2\pi \times 6$ MHz is the excited state decay rate. By varying $\delta$ from - 45 MHz to 45 MHz, we obtain the transmission spectra in Fig.~\ref{fig:2}(a). The OD in each channel is determined through theoretical fitting the transmission spectra. The storage efficiency $\eta$ is defined as $\eta=\frac{\int \mathrm{N_{out}} dt}{\int \mathrm{N_{in}} dt}$, where $\int \mathrm{N_{out(in)}} dt$ is the total photon number contained in the output (input) pulse. The residual magnetic field around the cold atomic ensemble is minimized by employing the microwave spectroscopy method, which is illustrated in detail in the supplementary material. The optical filter system in each signal channel consists of two Fabry-Perot etalon (with a bandwidth of 500 MHz) and a neutral filter (with a bandwidth of 10 nm). This configuration effectively isolates noise originating from the control beam and other scattered light, achieving an extinction ratio of 70 dB. The overall optical noise level in each channel is measured at 0.01 per second, which is significantly lower than that of the probe beam, indicating effective noise reduction.

\smallskip

\begin{backmatter}

\bmsection{Funding}
This work is supported by the National Natural Science Foundation of China (NSFC)(12104358, 12104361, 12304406, 12175168 and 92050103), Shaanxi Fundamental Science Research Project for Mathematics and Physics (22JSQ035), and Postdoctoral Fellowship Program of China Postdoctoral Science Foundation (CPSF) (GZC20232118).

%\bmsection{Acknowledgment}
%H.G., M.C., and P.Z. conceived the project. X.Y. and C.W. performed the experiments. X.Y., C.W., J.W., and Y.C. analyzed the data. X.Y., C.W., J.W., Y.C., H.C., R.D., S.Z., D.W., and F.L. contributed to the experimental setup and interpretation of the results. X.Y., C.W., H.G., and M.C. supervised the project. All authors contributed to writing the manuscript. X.Y. and C.W. contributed equally to the work.

\bmsection{Disclosures}
 The authors declare no competing financial interests.
 
\bmsection{Data Availability Statement}
Data underlying the results presented in this paper are not publicly available at this time but may be obtained from the authors upon reasonable request.

\bmsection{Supplement document} 
See Supplement for supporting content.

\end{backmatter}

%%%%%%%%%%%%%%%%%%%%%%% References %%%%%%%%%%%%%%%%%%%%%%%%%

%%%%%%%%%% If using BibTeX:
\bibliography{sample}

%%%%%%%%%% If preparing manually:
% \begin{thebibliography}{1}
% \newcommand{\enquote}[1]{``#1''}

% \bibitem{Zhang:14}
% Y.~Zhang, S.~Qiao, L.~Sun, Q.~W. Shi, W.~Huang, L.~Li, and Z.~Yang,
%   \enquote{Photoinduced active terahertz metamaterials with nanostructured
%   vanadium dioxide film deposited by sol-gel method,}
%   {\protect\JournalTitle{Optics Express}} \textbf{22}, 11070--11078 (2014).

% \bibitem{Optica}
% {Optica}, \enquote{{Optica Publishing Group},}
%   \url{http://www.opg.optica.org}.

% \bibitem{FORSTER2007}
% P.~Forster, V.~Ramaswamy, P.~Artaxo, T.~Bernsten, R.~Betts, D.~Fahey,
%   J.~Haywood, J.~Lean, D.~Lowe, G.~Myhre, J.~Nganga, R.~Prinn, G.~Raga,
%   M.~Schulz, and R.~V. Dorland, \enquote{Changes in atmospheric consituents and
%   in radiative forcing,} in \enquote{Climate Change 2007: The Physical Science
%   Basis. Contribution of Working Group 1 to the Fourth Assesment Report of
%   Intergovernmental Panel on Climate Change,}  S.~Solomon, D.~Qin, M.~Manning,
%   Z.~Chen, M.~Marquis, K.~B. Averyt, M.~Tignor, and H.~L. Miler, eds.
%   (Cambridge University Press, 2007).

% \end{thebibliography}

\end{document}